\documentclass[twocolumn,showpacs,pra,aps,superscriptaddress,floatfix]{revtex4-1}
\pdfoutput=1
\usepackage{graphicx}
\usepackage{braket} 
\usepackage{float}  				    
\usepackage{natbib}
\usepackage{bm}
\usepackage{amsmath}

\usepackage{color}
 \usepackage{soul}
 \usepackage{amsfonts}

 \newcommand{\uvec}[1]{\mathbf{\hat{#1}}}
 \definecolor{darkred}{rgb}{0.8,0.1,0.1}
 \definecolor{DARKRED}{rgb}{0.8,0.1,0.1}
 \definecolor{darkblue}{rgb}{0.1,0.1,0.7}
 \definecolor{bleudefrance}{rgb}{0.19, 0.55, 0.91}
 
 \DeclareMathOperator{\Tr}{Tr}
 \DeclareMathOperator{\e}{e}
 \DeclareMathOperator{\sech}{sech}

\usepackage{colordvi}

\definecolor{orange}{RGB}{255,127,0}

\begin{document}

\title{Topological Uhlmann phase transitions for a spin-$j$ particle in a magnetic field}

\author{D. Morachis Galindo}
\affiliation{Centro de Nanociencias y Nanotecnolog\'ia, Universidad
Nacional Aut\'onoma de M\'exico, Apartado Postal 14, 22800 Ensenada, B.C., M\'exico}%

\author{F. Rojas}
\affiliation{Centro de Nanociencias y Nanotecnolog\'ia, Universidad
Nacional Aut\'onoma de M\'exico, Apartado Postal 14, 22800 Ensenada, B.C., M\'exico}

\author{Jes\'us A. Maytorena}
\affiliation{Centro de Nanociencias y Nanotecnolog\'ia, Universidad
Nacional Aut\'onoma de M\'exico, Apartado Postal 14, 22800 Ensenada, B.C., M\'exico}

\date{\today}

\begin{abstract}
The generalization of the geometric phase to the realm of mixed states is known as 
 Uhlmann phase. Recently, applications of this concept to the field of topological insulators have been made and an experimental observation of a characteristic critical temperature at which the topological Uhlmann phase disappears has also been reported. Surprisingly, to our knowledge, the Uhlmann phase of such a paradigmatic system as the spin-$j$ particle in presence of a slowly rotating magnetic field has not been reported to date. Here we study the case of such a system in a thermal ensemble. We find that the Uhlmann phase is given by the argument of a complex valued second kind Chebyshev polynomial of order $2j$. Correspondingly, the Uhlmann phase displays $2j$ singularities, occurying at the roots of such polynomials which define critical temperatures at which the system undergoes topological order transitions. Appealing to the argument principle of complex analysis each topological order is characterized by a winding number, which happen to be $2j$ for the ground state and decrease by unity each time increasing temperature passes through a critical value. We hope this study encourages experimental verification of this phenomenon of thermal control of topological properties, as has already been done for the spin-$1/2$ particle.
\end{abstract}
\keywords{Geometric phase, Uhlmann, Topological, Mixed States, Spin-j Particle.}
\maketitle

\title{UhlmannJ}
\author{dmorachisgalindo }
\date{February 2021}


\maketitle

\section{Introduction}

 The emergence of the geometric phase in quantum physics has been a groundbreaking event\cite{Berry1984}. It has served as a tool to comprehend its fundamentals, as the spin statistics theorem\cite{BerryRobbinsSpinStatistics} or the Aharanov-Bohm effect\cite{Berry1984}. 
 In the field of condensed matter physics the geometric phase is a quantity that comes out in the theoretical description of the quantum Hall effect\cite{HallConductandTopoogical,Niu_1984}, the correct expression for the velocity of Bloch electrons\cite{Bohm}, ferromagnetism\cite{FerroPolaBerry,BerryPhaseonElectronic}, topological insulators\cite{ReviewTIs}, among other phenomena. It is also a central concept in holonomic quantum computation\cite{Sjoqvist}.
 
 Firstly proposed for adiabatic cyclic evolution\cite{Berry1984}, the geometric phase concept has already been broadened  to arbitrary evolution of a quantum state\cite{AAphase,MUKUNDAuno,MUKUNDA2} and is thus ubiquitous in quantum systems. Nevertheless, this approach to the geometric phase considers pure states only, while a more realistic description of quantum phenomena requires making use of the density matrix formalism. 
 
 The appropriate extension of the geometric phase to mixed states was developed by Uhlmann\cite{UhlmannOrig1,UhlmannOrig2} and is thus called the \textit{Uhlmann phase}.
 It has been theoretically studied in the context of $1D$ and $2D$ topological insulators\cite{ViyuelaOneDimensionalFermions,DensMatTIs,Viyuela_2015,PhysRevB.97.235141}, for example the 1D SSH model\cite{SSHmodel} and the Qi-Wu-Zhang 2D Chern insulator \cite{QWZ1}. A key feature of these systems is the appearance of a critical temperatures above which the Uhlmann phase vanishes regardless of the topological character of the system in the ground state. This temperature sets a regime of stability of the topological properties in such systems, and has already been experimentally observed in a superconducting qubit\cite{UhlmannObservation}, which gives the Uhlmann phase a higher ontological status. 
 Recently, a connection between a topological invariant (the Uhlmann number) and measurable physical quantities, like the dynamical conductivity was established through the linear response theory\cite{UhlmannLinearResponse}.
 
 For pure quantum systems, a paradigmatic example to illustrate the abstract notions of quantum holonomies is the spin-$j$ particle interacting with a slowly rotating magnetic field. In its original paper, Berry\cite{Berry1984} obtained the beautiful solid angle formula for the eigenstates of this system . More recently, a generalization of the solid angle formula for arbitrary spin-$j$ states has been found in terms of the state's Majorana constellation\cite{BerryPhaseMajorana,MajoranaAtomi} which also gives insight on their entanglement properties\cite{LiuBerryEntanglement}. From a more practical standpoint, the study of the geometric phase for a spin-$1/2$ is the workhorse to build 1-qubit holonomic quantum gates\cite{Sjoqvist,SjoqvistConceptualAspects}, with a possible extension to $SU(2)$ qudit gates for higher spins\cite{GenerationThreeLevelSU(2)}. This makes the study of the geometric phase for spin-$j$ particles of vital importance.
 
 Surprisingly, the problem of the Uhlmann phase of a spin-$j$ particle, even the $1/2$ case whose Hamiltonian models any two-level quantum system, has not been addressed yet, to our knowledge. Here, we calculated the Uhlmann phase of a spin-$j$ particle subjected to a slowly rotating magnetic field. We derived a compact analytical expression in terms of the argument of the complex valued second kind Chebyshev polynomials\cite{arfken,Gradshteyn} $U_{2j}(z)$ multiplied by the Pauli sign $(-1)^{2j}$. When the field describe a circle instead of a cone, the Uhlmann phase becomes topological with respect to temperature: the change from zero to $\pi$ (or viceversa) at certain critical temperatures related to the roots of $U_{2j}(z)$. Based on a theorem of complex analysis,\cite{VisualComplexFunctions,needham1998visual} we define the Chern-like Uhlmann numbers\cite{TwoDimensionalTwoCriticalViyuela} as winding numbers. For arbitrary direction of the external field, and as a function of temperature, the colored plot of the phase allows to identify
 visually the number of critical temperatures and the Uhlmann topological numbers for a given spin number $j$.
 
The paper is organized as follows. In section 2 we derive the Uhlmann phase for the spin-$j$ particle. In section 3 we show the emergence of multiple thermal topological transitions and obtain their corresponding Chern-like numbers numbers.  In section 4 we analyse the Uhlmann phase with respect to temperature and magnetic field's polar angle. Section 5 contains the conclusions. 

\section{Uhlmann phase for an arbitrary spin $\boldsymbol{J}$ in an external magnetic field}

The Uhlmann phase of a mixed quantum state is given by the expression\cite{Viyuela_2015,PhysRevB.97.235141}
\begin{align}\label{UhlmannPhase}
    \Phi_U &= \arg\left(\Tr\left[\hat{\rho}\,\mathcal{P}e^{\oint \hat{A}_U} \right] \right),
\end{align}

\noindent  where $\hat{\rho}$ is the system's density matrix with a spectral decomposition $\sum_kp_k\ket{k}\bra{k}$  and $\mathcal{P}$ is the path-ordering operator\cite{Bohm}. The Uhlmann connection $\hat{A}_U$ is given by \cite{PhysRevB.97.235141}
\begin{align}\label{UhlmannConnection}
    \hat{A}_U &= \sum\limits_{l,k}\frac{(\sqrt{p_l}-\sqrt{p_k})^2}{p_l+p_k}\bra{l}(d\ket{k})\ket{l}\bra{k},
\end{align}
\noindent where $d$ is the exterior derivative operator\cite{Bohm}. This equation is written in the density matrix eigenbasis, thus a parameter dependence on the eigenkets is to be understood.

The Hamiltonian of a spin-$j$ particle interacting with a magnetic field is expressed
\begin{align}
    \hat{H} &= B\uvec{n}\cdot\boldsymbol{{\hat{J}}},
\end{align}
\noindent where all physical constants which give raise to the interaction are taken into account in $B$. The vector operator $\boldsymbol{\hat{J}}$ has the components $(\hat{J}_x,\hat{J}_y,\hat{J}_z)$, where $\hat{J}_i$ are the usual angular momentum matrices of spin $j$\cite{Sakurai}. We will consider  the familiar fixed magnitude magnetic field that rotates along the $\uvec{z}$ axis at constant  frequency. The unit vector $\uvec{n}=(\sin\theta\cos\phi,\sin\theta\sin\phi,\cos\theta)$ is taken with fixed $\theta$, while 
$\phi$ changes during the evolution from $0$ to $2\pi$. For a thermal ensemble, the corresponding density matrix is written as
\begin{align}
    \hat{\rho} &= \frac{e^{-\beta B\uvec{n}\cdot\boldsymbol{\hat{J}}}}{Z}
\end{align}
\noindent where $\beta=1/k_BT$. The partition function $Z$ can be ignored in the calculation of the Uhlmann phase (\ref{UhlmannPhase}), since it is real and represents just a scaling factor of the complex number $\Tr[\hat{M}]$, where  $\hat{M}=\hat{\rho}\,\mathcal{P}\e^{\oint\hat{A}_U}$. The thermal basis in this case is given by a rotated $\ket{j,m}$ basis, which, in Euler angle representation is given by
\begin{align}
    \ket{j,m;\uvec{n}} &= e^{-i\phi\hat{J}_z}e^{-i\theta\hat{J}_y}e^{i\phi\hat{J}_z}\ket{j,m} \ .
\end{align}

\noindent The thermal occupation probabilities $p_m$ are readily seen to be $e^{-\beta B m}/Z$. 

 We now proceed to calculate the the Uhlmann connection for the problem at hand. The factor involving the thermal occupation probabilities in (\ref{UhlmannConnection})
 can be expressed as
 \begin{align}\label{Conexionfact1}
     \frac{(\sqrt{p_m}-\sqrt{p_{m'}})^2}{p_m+p_{m'}} &= 1-\sech[\beta B(m-m')/2],
 \end{align}
 \noindent while the factor involving the exterior derivative $d$ becomes 
 \begin{align}\label{Conexionfact2}
     \bra{m'}d\ket{m} = ie^{-i(m'-m)\phi}\sin\theta\bra{m'}\hat{J}_x\ket{m},
 \end{align}
 
 \noindent plus a negligible diagonal term. Inserting equation \eqref{Conexionfact1} and \eqref{Conexionfact2} into \eqref{UhlmannConnection} straightforwardly yields
 \begin{align}
     \hat{A}_U &= -i\eta\left(\hat{J}_z\sin\theta-e^{-i\phi\hat{J}_z}\hat{J}_xe^{i\phi\hat{J}_z}\cos\theta\right)d\phi,
\end{align}

\noindent where $\eta = \sin\theta[1-\sech(\beta B/2)]$. Calculation of the time ordered integral $\mathcal{P}\e^{\oint\hat{A}_U}$ is equivalent to solve a Schr\"odinger equation 
 \begin{align}\label{SchrodingerLike}
     i\frac{d}{d\phi}\hat{U} & = \hat{V}(\phi)\hat{U},
 \end{align}
 
 \noindent where the solution $\hat{U}$ is the is just the time ordered exponential and $\hat{V}(\phi)$ is $i\hat{A}_U$. Solving eq.\eqref{SchrodingerLike} for a closed loop followed by the direction of the external field results in\cite{Bohm}
 \begin{align}
     \mathcal{P}\e^{\oint\hat{A}_U} &= (-1)^{2j}e^{-i2\pi[(\eta\sin\theta-1)\hat{J}_z-\eta \cos\theta\hat{J}_x]}.
 \end{align}
 
 In order to obtain the Uhlmann phase, we need to take the trace of $\hat{M}$. Except for the lowest total angular momentum representations, its exact closed form of is very cumbersome. Also we would need to find the specific matrix $\hat{M}$ for every $j$, which is not very practical for our purposes. A way out of this pothole is noticing that the object we need to trace out belongs to the Lie group $SL(2,\mathbb{C})$ in the $(j,0)$ representation. Basic representation theory of this group\cite{Jeevanjee}  tells us that once the eigenvalues $(\lambda_+,\lambda_-)$ of the $(1/2,0)$ representation are known, the eigenvalues for higher $j$ are given by
 \begin{align}
     \lambda^{2j}, &\lambda^{2j-2}, \dots , \lambda^{-2j+2}, \lambda^{-2j}
\end{align}
with $\lambda=\lambda_+ = \lambda^{-1}_-$,
 where $\lambda_\pm$ are the eingenvalues obtained from $j=1/2$. The equality above follows from the property that the group has unit determinant. Having noted this, the trace of $\hat{M}$ is readily seen to be
\begin{align}
    \Tr\left[ \hat{\rho}\,\mathcal{P}e^{\oint\hat{A}_U}\right] &=
    \frac{\lambda^{2j+1}-\lambda^{-2j-1}}{\lambda-\lambda^{-1}}.
\end{align}

\noindent The calculation that remains is to obtain the exact form of the eigenvalue $\lambda$. Diagonalization of $\hat{M}$ in the $(1/2,0)$ representation yields 
\begin{align}
    \lambda = &z + \sqrt{z^2-1},
\end{align}

\noindent with  $z(\theta)$ being the complex variable
\begin{align}\label{variablezeta}
    z &= \cosh(\beta B/2)\cos(\pi C) - i\sinh(\beta B/2)\sin(\pi C)\frac{\cos\theta}{C}
    \end{align}
where $C(\theta) = \sqrt{1-\sin^2\theta\tanh^2(\beta B/2)}$. 
The function $z(\theta)$ defines a simple closed curve in the complex plane, this property being of fundamental importance in what follows. With all these ingredients the Uhlmann phase is readily obtained:
\begin{align}\label{Uhlmannspinj}
    \Phi_{U}^{(j)} & = \arg\left[(-1)^{2j}\frac{(z+\sqrt{z^2-1})^{2j+1}- (z-\sqrt{z^2-1})^{2j+1}}{2\sqrt{z^2-1}}\right]\nonumber\\
    &= \arg[(-1)^{2j}U_{2j}(z)],
\end{align}

\noindent where $U_{2j}(z)$ are the second kind Chebyshev polynomials\cite{Gradshteyn}; we will refer to them just as Chebyshev polynomials to simplify matters. Equation \eqref{Uhlmannspinj} is valid under cyclic adiabatic evolution and is exact in this regime. The upper index $(j)$ tells us that the Uhlmann phase is that of a spin-$j$ particle. Also, in the low temperature limit this equation reduces to the corresponding Berry phase\cite{Berry1984,Bohm,Viyuela_2015}.
We note that the nice compact form of result \eqref{Uhlmannspinj} is not simply a phase portrait of the Chebyshev polinomials in the whole complex plane, because the point $z(\theta)$ lies in a curve, with its shape depending on the parameter $\beta B$. It is interesting, however, the relation between the phase of polynomials $U_{2j}(z)$ and an observable phase of a quantum system.
Appearance of Chebyshev polynomials in the Uhlmann phase can be traced back to $\hat{H}$ pertaining to $\mathfrak{su}(2)$ algebra and the particle being in thermal equilibrium. This generates an element of $SL(2,\mathbb{C})$ via $\hat{\rho}\e^{\oint\hat{A}_U}$, whose eigenvalues can always be written as $v\pm\sqrt{v^2-1}$, $v\in\mathbb{C}$. It would be interesting to explore what kind of mathematical object the Uhlmann phase would be when considering a Hamiltonian that belongs to $\mathfrak{su}(n)$ for $n>2$. 

\begin{figure}
    \centering
    \includegraphics[scale=0.5]{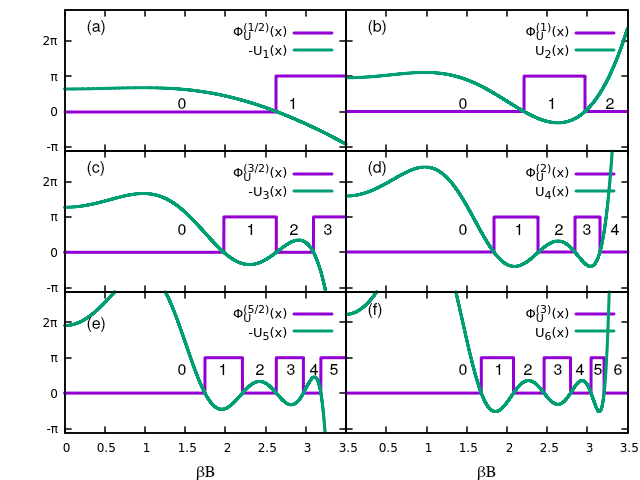}
    \caption{Uhlmann topological phases $\Phi^{(j)}_U$ and Chebyschev polynomials $(-1)^{2j}U_{2j}$ as functions of $\beta B$, for $ \theta=\pi/2$. The left column
    presents the phase for the half integer values $j=1/2, 3/2, 5/2$ ((a),(c), and (e), respectively). The right column displays the phase for the integer values $j=1,2,3$ ((b), (d), and (f)).
    The integers between the zeros of the polynomials indicate associated winding numbers, corresponding to Uhlmann numbers $n_U^{(j)}$ (see text in Sec.\,IIIB).}
    \label{figTjMultiple}
\end{figure}

\section{Topological Uhlmann phase transitions}

\subsection{Critical temperatures}

The Uhlmann phase just obtained is determined by the argument of the complex Chebyshev polynomials $U_{2j}(z)$. The function $U_{2j}$ have real roots only, $2j$ in number, lying in the open interval $(-1,1)$\cite{Gradshteyn}. The zeros of any polynomial $P_n(z)$ define points in the complex plane where its magnitude becomes zero, implying that its argument becomes undefined\cite{VisualComplexFunctions}. These points are referred to as phase singularities and are a general phenomenon of wave physics. In the field of optics,\cite{DENNIS2009293} for example, they allow to define optical vortices which have found a number of applications.

The Uhlmann phase displays $2j$ phase singularities.
The restriction on the variable $z(\theta)$ to acquire real values implies
that $\theta=\pi/2$, that is, the magnetic field should lie on the equator of
the sphere of directions. As a consequence, there are as many as $2j$ \textit{critical} temperatures $T^{(j)}_{c,k}$ ($k=1,\ldots,2j$) determined by
\begin{align}\label{temperaturasCriticas} \cosh(\beta_kB/2)\cos[\pi\sech(\beta_kB/2)] = \cos\left(\frac{k\pi}{2j+1}\right), 
\end{align}

\noindent where on the right hand side is the $k$-th root of the Chebyshev polynomial $U_{2j}(x)$, with $x=z(\theta=\pi/2)$.

Figure\,\ref{figTjMultiple} shows the Uhlmann phase as a function of $\beta B$  for $\theta=\pi/2$ and several values of $j$. Note that for half integer $j$ there is a negative sign multiplying $U_{2j}(x)$. It can be seen that 
$\Phi_U^{(j)}(\beta B)=\text{arg}[(-1)^{2j}U_{2j}(x)]$ is $0$ or $\pi$, and in this manner \textit{topological}. The topological transitions, between trivial and nontrivial phases,
occur at temperatures (or field magnitudes) such that the Chebyshev polynomials vanish, and the precise value, $0$ or $\pi$, of the phase $\Phi_U^{(j)}$ is determined by the sign of the polynomials times the Pauli sign $(-1)^{2j}$. Note that for very high temperatures ($\beta B\ll 1$) the Uhlmann phase vanishes, as expected for a system under thermal noise\cite{Viyuela_2015}. For very low temperatures ($\beta B\gg 1$), the phase $\Phi_U^{(j)}$ is either $\pi$ or zero for half integer and integer values of $j$, respectively. That is the expected behaviour because the Uhlmann phase of a thermal ensemble approaches the geometric phase of a pure system in its ground state as we approach zero temperature\cite{Viyuela_2015}. For example, a ground state spin-$j$ particle in a slowly rotating planar ($\theta=\pi/2$) magnetic field acquires a Berry phase\cite{Berry1984} $\gamma_{-j} = 2\pi j$,
consistent with the aforementioned. What is remarkable about this result is the emergence of many critical temperatures, distributed in a nonuniform way as $j$ varies.
There are $2j$ critical temperatures, some of which are at higher or lower values from that of the $j=1/2$ case. Thus, additional nontrivial topological phases appear at higher temperatures in comparison to the simplest spin one-half particle, and are in this sense more robust against thermal noise. On the other hand, the critical temperatures cannot reach arbitrary large values for high $j$, given the constraint imposed by eq.\eqref{temperaturasCriticas}, or equivalently, due to the fact that all the roots of $U_{2j}(z)$ lie in the interval $(-1,1)$.\cite{Gradshteyn}

Viyuela et al. predict\cite{TwoDimensionalTwoCriticalViyuela} the existence of two critical temperatures in a 2D topological insulator with high Chern numbers, suggesting the possibility of
purely thermal topological transitions.
Furthermore, the thermal topological phase transition for $j=1/2$ has already been confirmed experimentally\cite{UhlmannObservation} in a superconducting qubit. We regard this as a strong suggestion of the physical existence of multiple Uhlmann topological transitions for a spin-$j$ particle, and thus hope our results encourage experimental verification of this phenomenon.

\subsection{Topological Uhlmann numbers}

\begin{figure}
    \centering
    \includegraphics[scale=0.5]{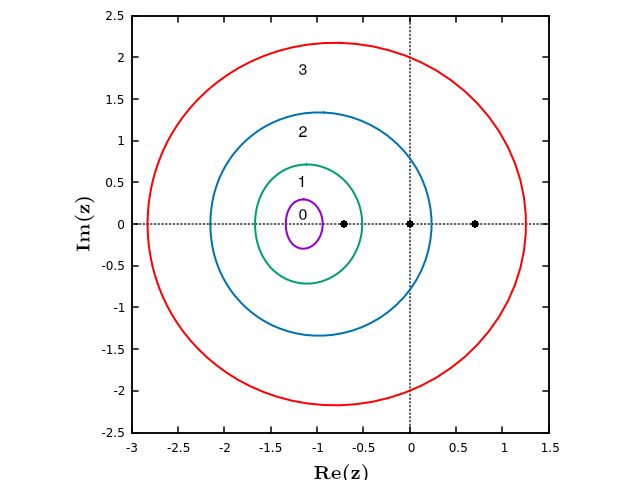}
    \caption{Argand diagram of $z(\theta)$ for several values of $\beta B$. Considering the case $j=3/2$, the points marked on the real axis are the roots of polynomial $U_3(z)$.
    At high enough temperatures the curve $z(\theta)$ (purple) does not enclose a single root. As temperature reduces the curve expands and encloses progressively the roots. The Uhlmann numbers are given by the number of roots inside $z(\theta)$.}
    \label{figGiros}
\end{figure}

The case $j=1/2$ is illustrative. There is only one topological
transition, occurring at $\beta B = 2\ln(2+\sqrt{3})$
(Fig.\,\ref{figTjMultiple}(a)). Viyuela et al.\cite{ViyuelaOneDimensionalFermions}report this single critical temperature\footnote{The extra factor of 2 arises from the definition $\hat{J}_i = \hat{\sigma}_i/2$ used for $j=1/2$.} for three representative 1D models of topological insulators and superconductors. At zero temperature, $T=0$, the ground state
of the system acquires a Berry phase of $\pi$ and a Chern number of $+1$. At finite temperature, for $T<T^{(1/2)}_{c,1}$ the same
phase is preserved, but above the critical temperature the Uhlmann topological phase becomes trivial. This is in sharp contrast to the zero temperature behavior.
A question that naturally arises at this point is about the invariants associated with the topological phases that occur for higher $j$. A look at Fig.\,\ref{figGiros} will give insight about writing the proper definition of them. The figure depicts the $z(\theta)$ curve \eqref{variablezeta} for four temperatures, where the dots mark the roots of $U_{3}(z(\theta))$. The smallest curve (purple) corresponds to the higher temperature while the largest (red) is for the lowest temperature. As the temperature decreases, the curve expands and progressively encloses the roots of $U_{3}$, whenever the temperature crosses a critical value $T^{(j)}_{c,k}$. The number of enclosed roots is zero at high temperatures, and ends up being $2j$ for low enough temperatures. This relates a topological property of $z(\theta)$, the number of roots enclosed, and the critical temperatures. 

According to the argument principle of complex analysis\cite{needham1998visual,VisualComplexFunctions}, if $z(\theta)$ encloses $k$ roots of $U_{2j}$, then the curve\footnote{If $z(\theta)$ is a simple closed curve in the interval $[0,\pi]$, so is $(-1)^{2j}U_{2j}(z(\theta))$} $U_{2j}(z(\theta))$ winds around the origin $k$ times, so the number of closed roots equals the winding number.
The winding number of the curve $(-1)^{2j}U_{2j}(z(\theta))$ tells us how many times its phase changes from $0$ to $2\pi$,\cite{Docarmo} but this phase is just $\Phi^{(j)}_U(\theta)$. This suggests the definition of Uhlmann numbers 
\begin{align}\label{numerosUhlmann}
    n^{(j)}_U(T) &= \frac{1}{2\pi}\int^{\pi}_0\frac{d\Phi^{(j)}_U(z(\theta))}{d\theta}d\theta,
\end{align}

\noindent to be the winding numbers of the $(-1)^{2j}U_{2j}(z(\theta))$ curve, for a temperature between two
succesive critical values. These integer numbers are the equivalent of the Chern numbers of pure 
states.\cite{BerryPhaseonElectronic} In fact, expression \eqref{numerosUhlmann} is consistent to that  
proposed as definition of Uhlmann numbers for two-dimensional topological insulators\cite{Viyuela_2015}. Here, we have followed
a more ad hoc path, motivated by the specific form of the Uhlmann
phase \eqref{Uhlmannspinj}, given in terms of polynomials.

\begin{figure}
    \centering
    \includegraphics[scale=0.5]{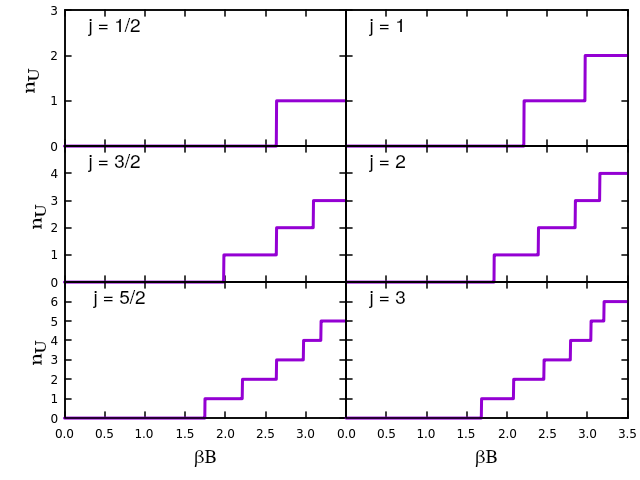}
    \caption{Uhlmann numbers $n_U$ as a function of temperature, for some values of the spin number $j$. The steps are located at a set of values of $\beta_k B$ which define critical temperatures $T^{(j)}_{c,k}$, with $k=1,\ldots,2j$.}
    \label{figChernUhlmannEscalera}
\end{figure}

Figure\,\ref{figChernUhlmannEscalera} shows the Uhlmann numbers for different values of $j$. The steps at which $n_U$ change by unity are located at the critical temperatures $T^{(j)}_{c,k}$. Note that the maximum value that $n_U$ takes on is $2j$, which equals the Chern number of a spin-$j$ particle in its ground state\cite{Bohm}. The figure illustrates how the appearance
of multiple critical temperatures makes possible transitions between nontrivial topological orders of the type $2j\to 2j-1\to 2j-2 \to\ldots\to 0$ for increasing temperature.

In the model of a 2D topological insulator which presents two critical temperatures\cite{TwoDimensionalTwoCriticalViyuela},
there are three topological  phases, one trivial with $n_U=0$ and
two nontrivial with $n_U=1$ and 2, which can be accessed by varying the temperature. The appearance of $2j+1$ distinct Uhlmann numbers in the spin-$j$ particle is a more dramatic example of a system with more than one nontrivial order. 

\section{Uhlmann geometric phases for arbitrary field direction}

\begin{figure}
    \centering
    \includegraphics[scale=0.5]{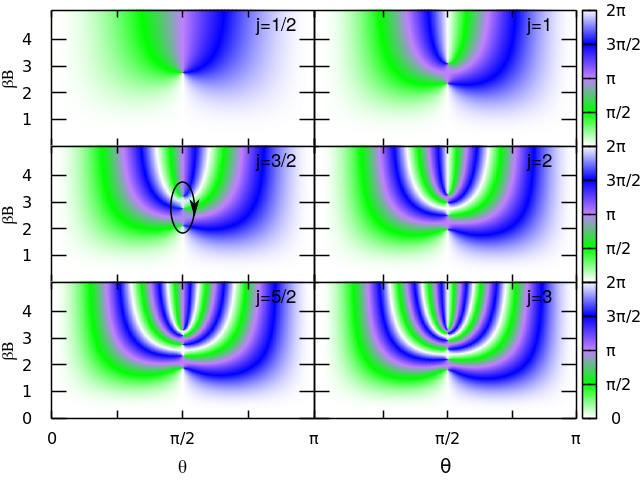}
    \caption{Uhlmann phases $\Phi^{(j)}_U(\theta,\beta B)$ for $j=1/2,1,\ldots,3$. There are $2j$ singularities along the
    $\theta=\pi/2$ line. The Uhlmann number
    characterizing a topological order
    can be obtained visually by encircling one or more singularities with a simple closed curve and counting the number of times an isochromatic line is repeated (see text).}
    \label{figUhlmannMult}
\end{figure}

In Fig.\,\ref{figTjMultiple} we show the temperature
dependence of the Uhlmann phase for $\theta=\pi/2$.
We now turn to analyse this dependence for directions in the
whole interval $0\leq\theta\leq\pi$. Fig.\,\ref{figUhlmannMult} shows a color map of $\Phi^{(j)}_U(\theta,\beta B)$ for distinct spin number $j$. In
each panel, $2j$ vortices can be distinguished along the line $\theta = \pi/2$, 
which correspond to the zeros of $U_{2j}(z)$, or equivalently, the critical temperatures. 
Note that for all cases the phase $\Phi^{(j)}_U(\theta,\beta B)\to 0$ for
high temperatures $\beta B\ll 1$, as expected. On the other hand, for very 
low temperatures $\beta B\gg 1$, the Uhlmann phase must converge to the Berry phase,\cite{Berry1984,Bohm} $\Phi^{(j)}_U(\theta,\beta B)\to 2j\pi(1-\cos\theta)$. 
For example, in the $j=1/2$ panel at low temperatures the sequence of colors as $\theta$ 
goes from zero to $\pi$ 
is that of the colorboxes on the right: the Uhlmann phase is $0$ for $\theta=0$ and 
increases up to $2\pi$ for $\theta = \pi$. Let us call  that sequence of colors a 
phase cycle. For panels with higher $j$ at low temperatures we see that the phase cycles 
appear $2j$ times. Lowering temperature, the number of phase cycles decrease by unity when crossing a critical temperature. A look to the $j=1$ panel illustrates this point. 
For low temperatures, there are two phase cycles as $\theta$ goes from $0$ to $\pi$. 
When increasing temperature above the first critical value $T^{(1)}_{c,1}$ 
the Uhlmann phase only traverses one phase cycle, and none of them above the 
second critical temperature. 

This behavior can also be illustrated in a similar way to that used to see the 
argument principle in action through
the colored phase portrait of a complex function.\cite{VisualComplexFunctions} 
To take a concrete example, consider a simple closed path encircling the 
three singularities of the phase in Fig.\,\ref{figUhlmannMult} for $j=3/2$, and follow the number of phase cycles occurring when
it is traversed. It can be seen that this number is exactly $2j$, the number of zeros enclosed, which is also the number of critical temperatures. The isochromatic lines (for instance the green ones) appear just $2j$ times. The number of cycles
diminish by one each time the path shrinks to leave out a zero,
where shrinks means to increase the temperature, in line with 
the geometrical interpretation of the Uhlmann numbers suggested by Fig.\,\ref{figGiros}. Thus, in our problem the Uhlmann numbers can also be obtained from the number of cycles displayed by the function $\Phi^{(2j)}_U(\theta,\beta B)$ in the vicinity of singularities.

\section{Conclusions}

In this paper we have studied the Uhlmann phase of a spin-$j$ particle interacting with a slowly varying magnetic field. We obtained a simple expression for that phase given by the argument of complex valued second kind Chebyshev polynomials $U_{2j}(z)$ multiplied by the Pauli sign $(-1)^{2j}$, the complex variable $z$ being a function of the direction of the external field and temperature.  As a consequence, $2j$ phase singularities appear which imply the possibility of topological phase transitions at $2j$ distinct critical temperatures. This is remarkably in contrast to the temperature dependence of the Uhlmann phase reported for topological insulators and superconductors.
Based on the principle argument of complex analysis, we derived
a proper topological invariant, the Uhlmann number, as a winding number associated to a topological order of the system, existing between two successive critical values of the temperature.
The Uhlmann number lie between $0$ and $2j$.

Our study suggests a purely thermal manipulation of topological transitions of a spin-$j$ particle.
This nontrivial effect has already been observed for the $j=1/2$ case and thus we hope this study encourages experimental verification of this phenomenon.

\section{Acknowledgements}

D.M.G. acknowledges support from Consejo Nacional de Ciencia y Tecnolog\'ia (M\'exico). The authors thank Ernesto Cota and Jorge Villavicencio for fruitful discussions.

\medskip

\bibliography{ref}

\end{document}